\documentclass[prb,twocolumn,showpacs,preprintnumbers,amsmath,amssymb,float,10pt,longbibliography]{revtex4-1}

\usepackage{subfigure}
\usepackage{graphicx}
\usepackage{amsmath}
\usepackage{physics}
\usepackage{hyperref}
\usepackage{epstopdf}
\usepackage{amsbsy}
\usepackage{placeins}
\usepackage{comment}
\usepackage{color}
\usepackage{dcolumn}
\usepackage{bm}
\usepackage{pstricks}

\newcommand{\cu}[1]{c_{#1 \uparrow}}
\newcommand{\cud}[1]{c_{#1 \uparrow}^\dagger}
\newcommand{\cd}[1]{c_{#1 \downarrow}}
\newcommand{\cdd}[1]{c_{#1 \downarrow}^\dagger}
\newcommand{\omegat}{\tilde{\omega}}
\newcommand{\ham}{\widetilde{\mathcal{H}}}

\newcommand{\fl}[1]{(#1)} 
\renewcommand\vec[1]{\ensuremath\boldsymbol{#1}}
\renewcommand{\Re}{\operatorname{Re}}
\renewcommand{\Im}{\operatorname{Im}}

\begin{document}

\title{Robustness of Quasiparticle Interference Test for Sign-changing  Gaps in Multiband Superconductors}
\author{Johannes H. J. Martiny,$^1$ Andreas Kreisel,$^2$ P. J. Hirschfeld,$^3$ and Brian M. Andersen$^1$}
\affiliation{$^1$Niels Bohr Institute, University of Copenhagen, Juliane Maries Vej 30, DK 2100 Copenhagen, Denmark}
\affiliation{$^2$Institut f\" ur Theoretische Physik, Universit\" at Leipzig, D-04103 Leipzig, Germany}
\affiliation{$^3$Department of Physics, University of Florida, Gainesville, FL 32611, USA}

\date{\today}

\begin{abstract}
Recently, a  test for a sign-changing gap function in a candidate multiband unconventional superconductor involving quasiparticle interference data was proposed.   The test  was based on the antisymmetric, Fourier transformed conductance maps integrated over a range of momenta $\bf q$ corresponding to interband processes, which was argued to display a  particular resonant form,  provided the gaps changed sign between the Fermi surface sheets connected by $\bf q$.  The calculation was performed for a single impurity, however, raising the question of how robust this measure is as a test of sign-changing pairing in a realistic system with many impurities.   Here we reproduce the results of the previous work within a model with two distinct Fermi surface sheets, and show explicitly that the previous result, while exact for a single nonmagnetic scatterer and also  in the limit of a dense set of random impurities, can be difficult to implement for a few dilute impurities.  In this case, however, appropriate isolation of a single impurity is sufficient to recover the expected result, allowing a robust statement about the gap signs to be made.
\end{abstract}
\maketitle
\section{Introduction}
The most fundamental question that can be posed about a newly discovered superconductor is why conduction electrons pair.  In particular, it is important to know whether the pairing mechanism is unconventional, i.e. based on an effective attraction that arises from the Coulomb interaction rather than from the  exchange of phonons as in a traditional BCS system.  While this question is difficult to answer directly, a good hint is often provided by whether the gap changes sign over the Fermi surface of the candidate material, since such a  state is generically produced by repulsive interactions.  Even this indirect type of proof has been particularly  vexing to establish in the iron-based superconductors (FeSC)\cite{HirschfeldCRAS,Stewart11,Chubukov_AnnRev12}, however, since many of them have rather isotropic gaps, and the sign-change may occur between different Fermi surface sheets, leading to a quasiparticle spectrum without zero-energy states,  very similar to that which would occur were the system to have the same gaps without a sign change.

Recently, an apparently simple method to  detect the existence of a gap sign change {\it qualitatively}  was proposed by Hirschfeld, Altenfeld, Eremin and Mazin (HAEM) \cite{HAEM_2015}.  They pointed out that when quasiparticle interference (QPI)  data are collected in a scanning tunneling spectroscopy (STS) experiment, certain QPI peaks correspond to interband scattering, and are coherently enhanced, with a characteristic bias dependence if the gap function changes sign between the two bands.  The proposal was similar in spirit  to an earlier one by the authors of Ref. \onlinecite{Chi_2014}, who however employed a  heuristic expression for the QPI amplitude based on BCS coherence factors and a Fermi Golden Rule argument.  Ref. \onlinecite{HAEM_2015} derived the correct expression for the observable of interest, the antisymmetric  momentum-integrated tunneling conductance, and showed in a simple model band structure how to distinguish between a sign changing $s$-wave ($s_\pm$) and non-sign-changing ($s_{++}$) state.  

The approach of Ref. \onlinecite{HAEM_2015} was simplified in certain ways that may be legitimately questioned, however.  The main issue is that the QPI signal was calculated for a single impurity.    In the presence of $N$ impurities at random positions, the Friedel oscillations emanating from these scatterers interfere.  For sufficiently many impurities, the random phases of the waves average to zero as 1/$\sqrt{N}$,\cite{Zhu_2014} and the QPI results should resemble the single-impurity results closely, but the concentration for which this occurs is difficult to pinpoint.  We therefore study the problem of many impurities in a multiband superconductor, in order to understand more precisely under which circumstances the HAEM result may be expected to hold, and to investigate practical ways of using the vast quantities of STS data available to draw the cleanest conclusions regarding the gap signs.  We show that the HAEM result is indeed obscured for a dilute impurity system, but also that a clear signal can generally be recovered by properly windowing an isolated impurity that occurs in such a context, a technique already successfully employed on the Fe-based superconductor FeSe\cite{OSP1}. We further investigate and discuss the effect of stronger and magnetic impurity potentials.

Finally, we discuss how to apply the HAEM method to non-$s$-wave pair states in multiband systems.  An interesting distinction arises in the case of $d$-wave states in systems without hole pockets, proposed e.g. for monolayer FeSe on SrTiO$_3$ and various FeSe intercalates (for a discussion see Ref. \onlinecite{HirschfeldCRAS}).  Since the gap sizes on the two electron pockets  are identical by symmetry, the simplest nodeless $d$-wave case yields a sharp, antisymmetrized momentum-integrated conductance peak whose width is due only to on-pocket anisotropy and thermal broadening.  Such a signal should be easy to distinguish from that arising from both $s_\pm$ and $s_{++}$ gap structures.   

\section{Method}
\subsection{Model}
\begin{figure}[tb]
\includegraphics[width=\linewidth]{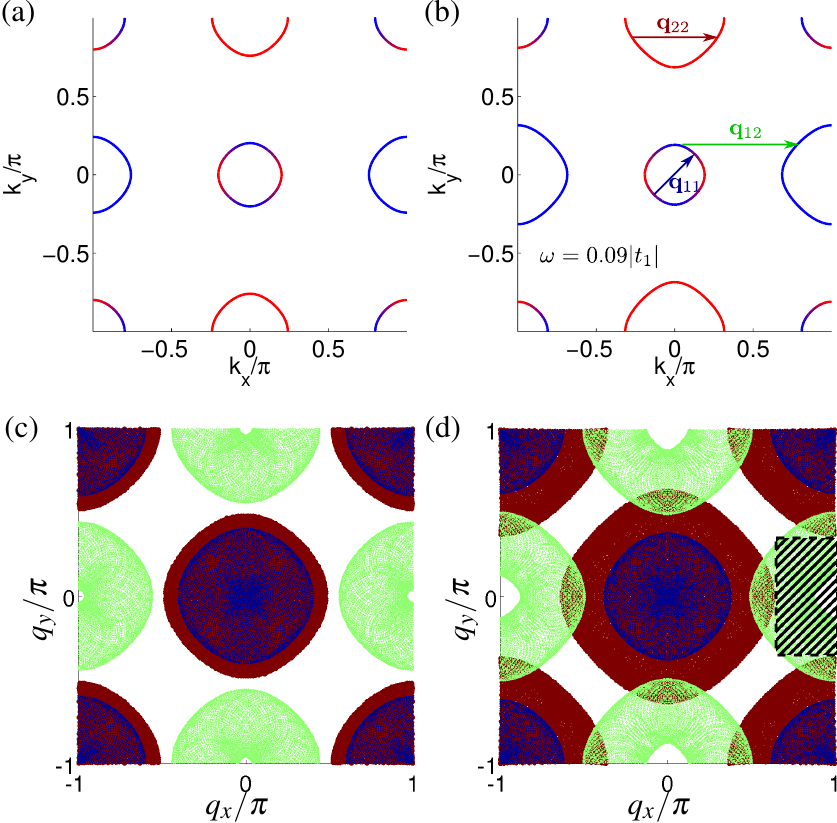}
\caption{(a) Fermi surface of the two-orbital model together with the orbital character (blue/red). A constant energy surface at $\omega=0.09|t_1|$ is plotted in (b) together with examples of intraband scattering vectors ${\mathbf q}_{11}$ and ${\mathbf q}_{22}$ which involve scattering without sign change of the order parameter, and interband scattering vectors ${\mathbf q}_{12}$ which for the $s_\pm$ order parameter involves scattering with sign change. To visualize the expected area in q-space of these scattering vectors we show in (c) a joint density of states (JDOS) visualization at $\omega=0$ distinguishing the different scattering vectors by the same color code as in (b), and in (d) the same at $\omega=0.09|t_1|$ where the scattering vectors start to significantly overlap  in q-space. In (d) the area $\mathcal C$ is marked in one quadrant as a shaded area only containing interband scattering vectors ${\mathbf q}_{12}$.
}
\label{fig:JDOS}
\end{figure}

Our starting point is the band structure model developed for LaOFeAs by Qi {\it et al.}\cite{Qi2008}, here presented in terms of the spinor $\psi_{k\sigma}^\dagger =(c^\dagger_{xz,\sigma},c^\dagger_{yz,\sigma})$ describing the "$d_{xz}$" and "$d_{yz}$" orbitals of a square Fe lattice, 
\begin{align}
\mathcal{H}_0 &= \sum_{k\sigma} \psi^\dagger_{k\sigma}\left[(\epsilon_{+}(k)-\mu)\tau_0+\epsilon_{-}(k)\tau_3+\epsilon_{xy}(k)\tau_1 \right] \psi_{k\sigma},
\end{align}
where $\tau_0$ is the identity matrix and $\tau_i$ denote the Pauli matrices. The three terms in the Hamiltonian are described by the dispersions
\begin{align}
\epsilon_{+}(k) = &-(t_1+t_2)[\cos(k_x)+\cos(k_y)]\nonumber \\
 	&-4t_3 \cos(k_x)\cos(k_y),\\
\epsilon_{-}(k) = &-(t_1-t_2)\left[\cos(k_x)-\cos(k_y)\right],\\
\epsilon_{xy}(k)= &-4 t_4 \sin(k_x)\sin(k_y).
\end{align}
This two-orbital model exhibits a Fermi surface with electron and hole pockets at the $X,Y$ and $\Gamma, M$ points, respectively. Clearly the band structure of real systems is considerably more complex, but for the purposes of this paper, the model above suffices. We choose hopping parameters $t_i$ such that the size of pockets are reduced from the original Qi {\it et al.} result, fixing $t_1= -1, t_2 = 1$, and $t_3 = t_4 = -0.63$, see Fig.~\ref{fig:JDOS} \fl{a} for the Fermi surface. The small pocket sizes enable the separation of intra-  and inter-band (e.g. $\vec{q}_{12}$ in Fig.~\ref{fig:JDOS} \fl{b}) nesting vectors on the equal energy contours at low energy, which makes the HAEM procedure more straightforward, see Fig.~\ref{fig:JDOS} \fl{b}.\\ 

\subsection{Self-consistent Chebyshev-BdG}
We include superconductivity (SC) at the mean field level 
\begin{align}
\mathcal{H}_{SC} &=  -\sum_{k, \mu} \Delta_k^\mu c_{k \mu \uparrow}^\dagger c_{-k \mu \downarrow}^\dagger+ \text{H.c.},
\end{align}
with $\Delta_k^\mu = V \ev{c_{-k \mu \downarrow}c_{k \mu \uparrow}}$. We consider two distinct model gap structures  relevant for FeSC,
\begin{align}
\Delta_k^{++} &= \Delta_0^{++}+ \Delta_0^{ext}(\cos(k_x)+\cos(k_y)), \\
\Delta_k^{+-} &= \Delta_0^{+-}\cos(k_x)\cos(k_y)+ \Delta_0^{ext}(\cos(k_x)+\cos(k_y)).
\end{align}
Here, the first term in the second expression has line nodes in the regions of the BZ separating the electron and hole pockets, i.e. the gaps on these pockets have opposite signs.
 Distinct magnitudes of the gaps on the electron and hole pockets for both gap structures is obtained by including a sub-dominant extended $s$-wave term, 
 with $\Delta^0_{ext}< \{\Delta_0^{+-},\Delta_0^{++}\}$, chosen small enough to preserve the sign change for the $s^{+-}$ gap structure. \\
 
For selfconsistent calculations we generate the above gap structures by including the corresponding real space terms
\begin{align}
\mathcal{H}_{SC}^{++}&= -\sum_{ i,\mu } \Delta_{i, \mu}^{++} c_{i\mu \uparrow }^\dagger c_{i\mu \downarrow}^\dagger  - \sum_{\langle ij \rangle,\mu } \Delta_{ij, \mu}^{++} c_{i\mu \uparrow }^\dagger c_{j\mu \downarrow}^\dagger + \text{H.c.}, \\
\mathcal{H}_{SC}^{+-} &= - \sum_{\langle \langle il \rangle \rangle,\mu } \Delta_{il, \mu}^{+-} c_{i\mu \uparrow }^\dagger c_{l\mu \downarrow}^\dagger - \sum_{\langle ij \rangle,\mu } \Delta_{ij, \mu}^{+-} c_{i\mu \uparrow }^\dagger c_{j\mu \downarrow}^\dagger + \text{H.c.},
\end{align}
with $\langle \rangle$, $\langle \langle \rangle \rangle$ indicating summation over nearest- and next-nearest neighbor sites in the square lattice, and $\Delta_{ij,\mu}^{\alpha}= V^{\alpha}\ev{\cd{j\mu} \cu{i\mu}}$. The pairing potentials are chosen as $\{V^{++}_i, V^{++}_{\langle ij \rangle} \}/|t_1| = \{0.73, 0.26 \}$ and $ \{V^{+-}_{\langle \langle il \rangle \rangle }, V^{+-}_{\langle ij \rangle} \}/|t_1| = \{0.46,0.31\}$.

We write the real space Hamiltonian in terms of the Nambu vector 
$\mathbf{c}_{i\mu}^\dagger = (c_{i\mu \uparrow}^\dagger,  c_{i\mu \downarrow})$, yielding
\begingroup
\renewcommand*{\arraystretch}{1.5}
\begin{align}
\mathcal{H} &= \sum_{ij,\mu \nu} \mathbf{c}_{i\mu}^\dagger H_{ij}^{\mu \nu} \mathbf{c}_{j\nu},  \\
H_{ij}^{\mu \nu} &=\begin{pmatrix}
\tilde{t}_{ij}^{\mu \nu} & \overline{\Delta}_{ij}^{\mu} \\
\Delta_{ij}^{\mu} & -\tilde{t}_{ij}^{\mu \nu} \\
\end{pmatrix},
\end{align}
\endgroup
where we have set $\tilde{t}_{ij}^{\mu \nu}=  (t_{ij}^{\mu \nu}-\delta_{ij}\delta_{\mu \nu}\mu_0)$. 
Including all sites and orbitals in the Nambu vector, the above matrix $H$ is of dimension $N_x^2 \times 2 \times 2$.\\

With the Hamiltonian established, we initially perform self-consistent calculations on a square lattice of dimension $101 \times 101$, and include a single central impurity, which we model as a delta function repulsive potential
\begin{align}
\mathcal{H}_{imp} &= V_{\text{imp}} \sum_{\mu, \sigma} c^{\dagger}_{i' \mu \sigma } c_{i' \mu \sigma},
\end{align} 
with $V_{\text{imp}} = 0.01$ the impurity potential and $i' $ indicating the central impurity site.  
A large lattice size, and hence good k-space resolution, is required for the calculations, which we obtain by using the Chebyshev-Bogoliubov-de Gennes (CBdG)  method\cite{Weisse_KPM,Covaci2010}. 
This method removes the computationally complex diagonalization of the full Hamiltonian, by  expanding instead the Greens function in a series of Chebyshev polynomials. 
These functions are defined on the interval $\omega \in(-1,1)$, which we accommodate by rescaling the Hamiltonian
\begin{align}
\ham = (\mathcal{H}-b)/a,
\end{align}
in terms of the parameters $b=(E_{max}+E_{min})/2$ and $a=(E_{max}-E_{min})/\eta$, where $\eta$ is a small parameter introduced to avoid divergence at the edges of the interval, and the rescaled Hamiltonian has eigenvalues in the interval $\tilde{\omega} \in (-1,1)$. The extremal eigenvalues can be estimated from smaller system size calculations, and no change in the mean fields is found from overestimation of these values, as was also found in Ref. \onlinecite{Nagai2012}.\\

We calculate the four components of the Nambu Greens function
\begin{align}
\bar{G}_{ij}(\omegat) &= 
\lim_{\eta \to 0}\mel{0}{\begin{pmatrix}\cu{i} \\ \cdd{i} \end{pmatrix} \frac{1}{\omegat + i \eta - \ham} (\cud{j} \cd{j})}{0},
\end{align}
e.g. the normal Greens function
\begin{align}
\bar{G}_{ij}^{11}(\omegat) &= \lim_{\eta \to 0}\mel{\cu{i}}{\frac{1}{\omegat + i \eta - \ham}}{\cud{j}}  \\
&= \frac{-2i}{\sqrt{1-\omegat^2}}\sum_{n=0}^N a_n^{11}(i,j)\exp(-in \arccos(\omegat)) \nonumber,
\end{align}
with $\ket*{\cud{j}} = \cud{j} \ket{0}$, and coefficients for the normal and anomalous components 
\begin{align}
a_n^{11}(i,j) &= \frac{1}{1+\delta_{0,n}} \mel{\cu{i}}{T_n(\ham)}{\cud{j}}, \\
a_n^{12}(i,j) &= \mel{\cud{i}}{T_n(\ham)}{\cud{j}}^*,   
\end{align}
where $T_n$ is the $n$th Chebyshev polynomial of the first kind. 
These expansion coefficients are obtained using the recursion relation of the $T_n$. Setting $\ket{j_n} = T_n(\ham) \ket{\cud{j}}$, we find
\begin{subequations}
\begin{align}
\ket{j_0} &= \ket*{\cud{j}} \\
\ket{j_1} &= \ham \ket*{\cud{j}} \\
\ket{j_{n+1}} &= 2\ham \ket{j_n}-\ket{j_{n-1}},
\end{align}
\end{subequations}
and the full expansion coefficients are obtained by $a_n^{11}(i,j) = \braket*{\cu{i}}{j_n}, a_n^{12}(i,j) = \braket*{\cud{i}}{j_n} $. Once we have computed the vectors $\ket{j_n}$ the coefficients for all $i$ are then readily available. \\

We then calculate the local density of states (LDOS), electron density and SC mean fields in terms of the above Greens functions
\begin{align}
A^{\uparrow (\downarrow)}_{i}(\tilde{\omega}) &= -\frac{1}{\pi} \Im G^{11(22)}_{i}(\tilde{\omega}), \\
n_i &= -\frac{1}{\pi}\int_{-1}^1 d\tilde{\omega} \left[ \Im G^{11}_{ii}(\tilde{\omega})+\Im G^{22}_{ii}(\tilde{\omega}) \right]f(\tilde{\omega}), \\
\Delta_{ij} &=-\frac{V}{2\pi} \int_{-1}^1 d\omegat~ \Im G^{12}_{ij}(\tilde{\omega}) [1-2f(\omegat)],
\end{align}
with $f(\omegat)$ the Fermi-Dirac distribution. These integrals are efficiently evaluated using Chebyshev-Gauss quadrature as discussed in Ref. \onlinecite{Weisse_KPM}. A small broadening of $1$meV is included in the spectral function by applying the Lorentz kernel, modifying the expansion coefficients\cite{Weisse_KPM}.

\begin{figure}[t]
\centering
\subfigure{\includegraphics[width = \linewidth]{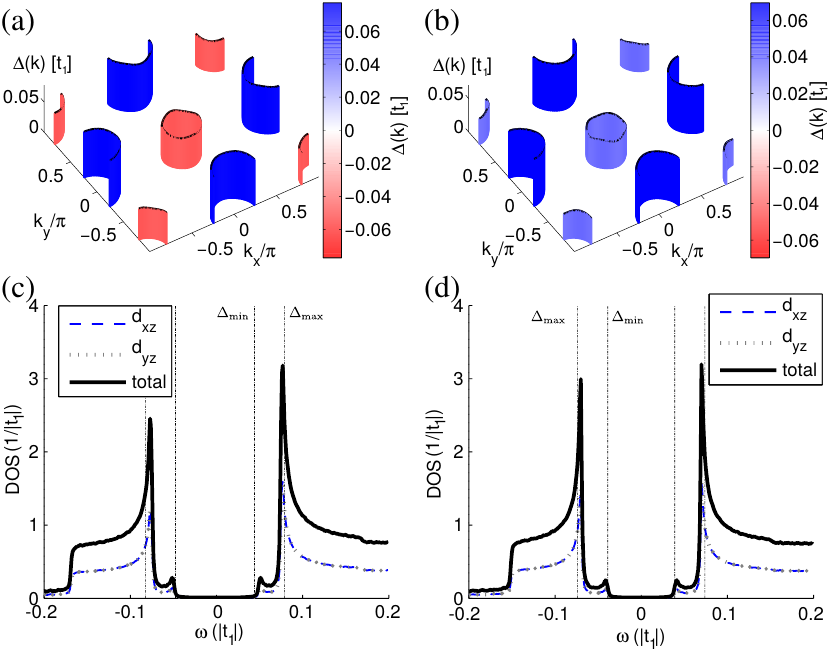}}
\caption{
\fl{a-b} Superconducting gap $\Delta(k)$ on the Fermi surface and the corresponding high resolution DOS \fl{c-d} obtained from a k-space BdG calculation of the spectral function, treating the CBdG converged mean fields as inputs. Panels (a,c) and (b,d) refer to $s_\pm$ and $s_{++}$, respectively. }
\label{fig:Tuning}
\end{figure}

The converged gap structures are plotted on the Fermi surface in Fig.~\ref{fig:Tuning} \fl{a-b}, demonstrating slightly anisotropic gaps of the electron and hole pockets of different magnitudes. Figure~\ref{fig:Tuning} \fl{c-d} shows the high resolution DOS obtained from a k-space BdG calculation using the converged CBdG mean fields as inputs. A clear two-gap structure with the gaps on both electron and hole pockets of similar magnitudes for $s_{++}$ and $s_{\pm}$ is obtained by fine-tuning the pairing potentials. Thus we have obtained two systems with indistinguishable DOS but distinct signs on the gaps between the electron and hole-pockets, enabling direct comparison between the HAEM QPI curves in these systems.\\    

\subsection{Bogoliubov QPI signal}
The CBdG method yields the LDOS $\rho(\vec{r},\omega)$ directly. Subtracting the constant LDOS in the homogeneous system isolates the impurity induced change, and we can then define the FT-LDOS
\begin{align}
\delta \rho(\vec{q},\omega) &= \frac{1}{N_x^2}\sum_{\vec{r}} e^{-i\vec{q} \cdot \vec{r} }\delta \rho(\vec{r},\omega).
\end{align}

The sharp peaks corresponding to the interpocket signal reside at large $\vec{q}$ and we thus define the interband signal as 
\begin{align}
\delta\rho_{inter}(\omega) &= \sum_{\vec{q} \in \mathcal{C}} \delta \rho(\vec{q},\omega)
\end{align}
with $\mathcal{C}$ a $\vec{q}$ region containing purely interband scattering vectors (see Fig. \ref{fig:JDOS}). The size of this integration area is set by considering the joint density of states (JDOS) at $\omega < \Delta_{max}$, as shown in Fig.~\ref{fig:JDOS} \fl{c-d}. This quantity has central weight corresponding to intraband contributions, weight at the $M$ points corresponding to hole-hole (and electron-electron) pocket scattering, and finally weight at the $X,Y$ points corresponding to electron-hole pocket scattering. Since we wish to consider interband scattering with equal/opposite gap signs, we isolate this final region as illustrated by the shaded black box in Fig.~\ref{fig:JDOS} \fl{d}, as well as the equivalent boxes at $(-\pi,0)$ and $(0,\pm \pi)$ (not shown).
Following Ref. \onlinecite{HAEM_2015,OSP1}, we then define the antisymmetrized QPI signal by
\begin{align}
\delta \rho_{inter}^{-}(\omega) &= \Re(\delta \rho_{inter}(\omega)) - \Re(\delta \rho_{inter}(-\omega)).
\label{eq_anti_qpi}
\end{align}

Experimental FT-STM measurements are usually obtained with the application of a window function prior to the FT to avoid spectral leaking from the differential conductance discontinuity at the field of view (FoV) edge. To enable comparisons between our periodic lattice and the experimental FoV, we apply the window function

\begin{align}
W(x,y) &= \cos(x\pi/I)\cos(y\pi/I), \label{eq:Window}
\end{align}
with $I$ the FoV dimension, chosen initially as the full lattice length. 

\subsection{T-matrix approach}

{\renewcommand{\hat}{}
\renewcommand{\mathbf}{\vec}
To test the convergence for a single impurity in a large system, we also perform T-matrix calculations of the impurity induced LDOS change.
For this purpose, we construct the matrix Greens function
 \begin{equation}
  {\hat G}_0(\mathbf{k},\omega)=(\omega -\mathcal{H}(\mathbf{k})+i\eta)^{-1},
 \end{equation}
and calculate the local Greens function ${\hat G}_0(\omega)=\sum_{\mathbf{k}}{\hat G}_0(\mathbf{k},\omega)$.
Here $\mathcal{H}=(\mathcal{H}_0,\Delta(\mathbf{k});\Delta(\mathbf{k})^*,-\mathcal{H}_0)$ is the Nambu Hamiltonian with the superconducting order parameter $\Delta(\mathbf{k})$ in orbital space.
Together with the impurity potential 
 we can calculate the T-matrix
\begin{equation}
 {\hat T}(\omega)=[1-{\hat V}_{\text{imp}}\tau_3{\hat G}(\omega)]^{-1} {\hat V}_{\text{imp}}\tau_3\,,
\end{equation}
with $\tau_3=(1,0;0,-1)$.
The full Greens function in the presence of the impurity within the T-matrix approach is then given by
\begin{equation}
  {\hat G}(\mathbf{k},\mathbf{k'} ,\omega)= {\hat G}_0(\mathbf{k},\omega)+{\hat G}_0(\mathbf{k},\omega){\hat T}(\omega){\hat G}_0(\mathbf{k'},\omega)\,.
\end{equation}
Noting that the T-matrix is momentum independent for a pure on-site potential scatterer,
we can calculate the impurity contribution to the Green function
\begin{equation}
  \delta {\hat G}(\mathbf{q} ,\omega)= \sum_{\mathbf{k}}{\hat G}_0(\mathbf{k},\omega){\hat T}(\omega){\hat G}_0(\mathbf{k+q},\omega)\,,
\end{equation}
such that the contribution to the LDOS is given by
\begin{equation}
 \delta \rho(\mathbf{q} ,\omega)= -\frac{2}{\pi} \mathop{\text{Im\,Tr'}} \delta {\hat G}(\mathbf{q} ,\omega),
\end{equation}
where $\mathop{\text{Tr'}}$ is the trace over the first half (particle part) of the Greens function only and the factor 2 accounts for the spin-degeneracy. This is the same quantity as obtained by the CBdG method.
For the calculations including a magnetic contribution $J$ to the impurity potential, we use the impurity Hamiltonian
\begin{align}
\mathcal{H}_{\text{imp}} &=  \sum_{\mu, \sigma=\pm} (V_{\text{imp}}-\sigma J)c^{\dagger}_{i' \mu \sigma } c_{i' \mu \sigma},\label{eq_mag_V}
\end{align}
and then calculate the spin-summed LDOS via
\begin{equation}
 \delta \rho(\mathbf{q} ,\omega)= -\frac{1}{\pi} [\mathop{\text{Im\,Tr'}} \delta {\hat G}(\mathbf{q} ,\omega)-\mathop{\text{Im\,Tr''}} \delta {\hat G}(\mathbf{q} ,-\omega)],
\end{equation}
where  $\mathop{\text{Tr''}}$ is the trace over the second half (hole part) of the Greens function only.}

\newpage
\section{Results}

\begin{figure}[t]
\centering
\subfigure{
\includegraphics[width=1 \linewidth]{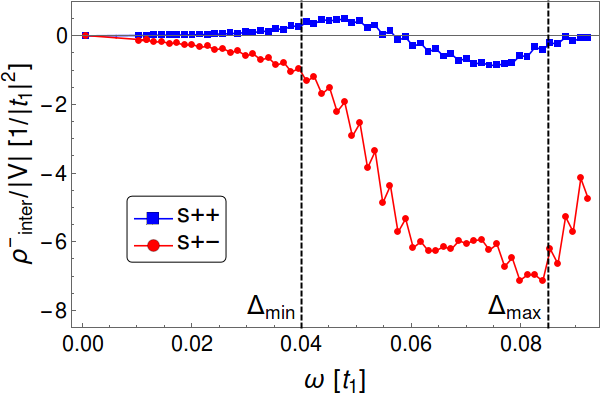}}
\caption{Integrated antisymmetrized FT-LDOS ($\delta \rho_{inter}^{-}(\omega)$) for $s_{++}$ (blue curve) and $s_{\pm}$ (red curve) gap structures, shown here normalized by the impurity potential $V_{imp} /|t_1|=0.01$. The system size is $N_x=101$. The result from Ref.~\onlinecite{HAEM_2015} is reproduced, with clear distinction between the different gap structures.  Dashed vertical lines are inner and outer gap edges.}
\label{fig:RhoInter1}
\end{figure}

The result of integrating the antisymmetrized interband signal from the CBdG calculation produces $\delta \rho_{inter}^{-}(\omega)$ which is shown in Fig.~\ref{fig:RhoInter1} as a function of energy $\omega$ for the case of a single FoV-centered impurity potential. The region of interest is indicated by the dashed vertical lines denoting the two gap magnitudes, i.e. the inner and outer gap edges seen in Fig.~\ref{fig:Tuning} \fl{c-d}. We find a clear sign crossing in the $s_{++}$ signal in the region of interest, while the $s_{\pm}$ signal is larger and sign preserving. We stress that this central impurity result for distinguishing between $s_{\pm}$ and $s_{++}$ is independent of the calculation details, i.e. variations in the impurity strength, integration area or spectral broadening leaves the qualitative result invariant until impurity bound states are formed\cite{HAEM_2015}. \\

\begin{figure}[t]
\includegraphics[width=\linewidth]{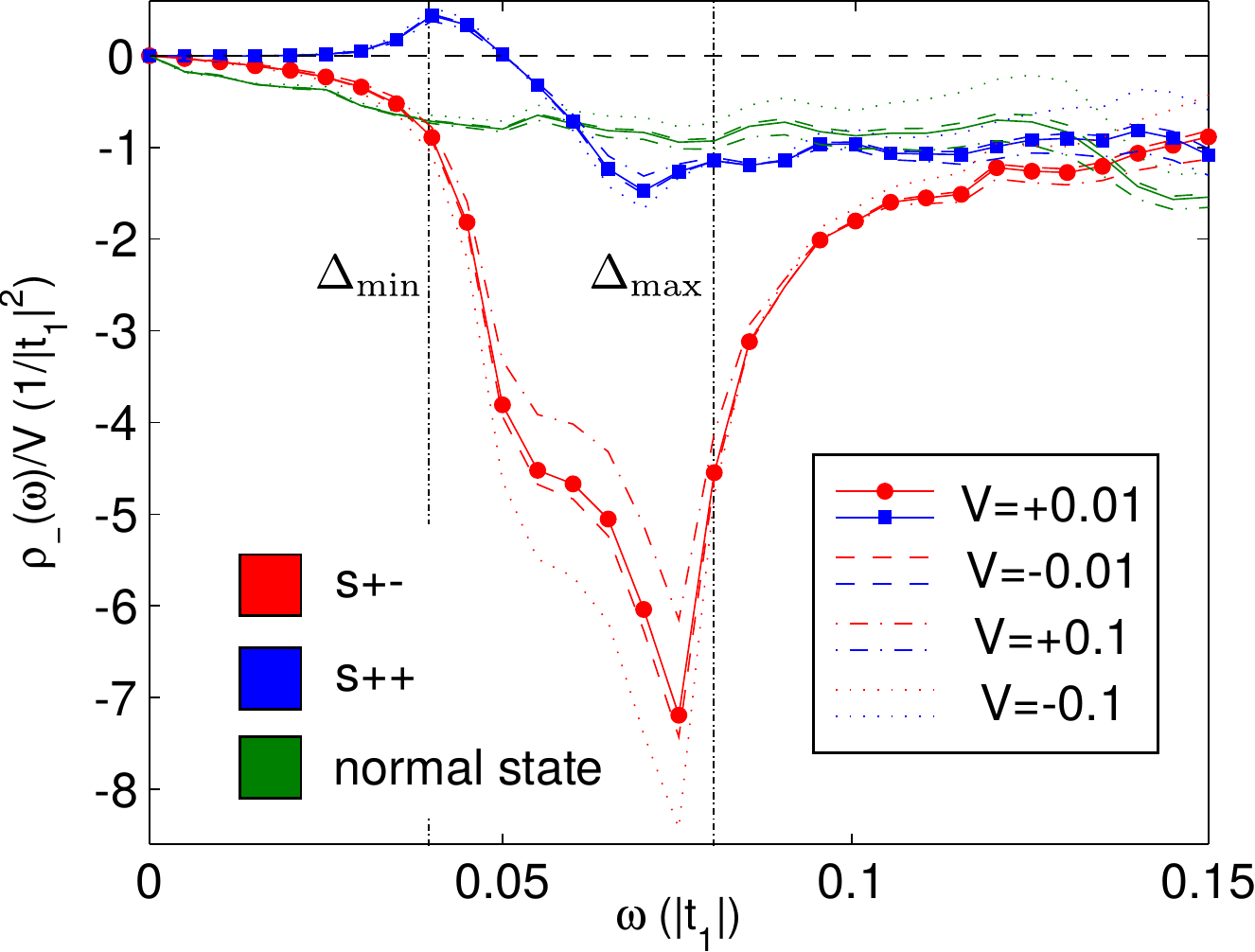}
\caption{Results of a T-matrix calculation for the antisymmetric QPI signal for various impurity potentials $V=\{\pm 0.01,\pm 0.1\}|t_1|$. The trivial sign and prefactor from the impurity potential was removed by dividing by the potential. In this way the difference between the sign-changing order parameter ($s_{\pm}$ red) and the non-sign changing order parameter ($s_{++}$ blue) becomes apparent, and is independent of the choice of the impurity potential. The same calculation carried out in the normal state (green) reveals that only a weak signal from the particle-hole asymmetry of the band structure is obtained.}
\label{fig:AK1}
\end{figure}

\begin{figure}[t]
\includegraphics[width=\linewidth]{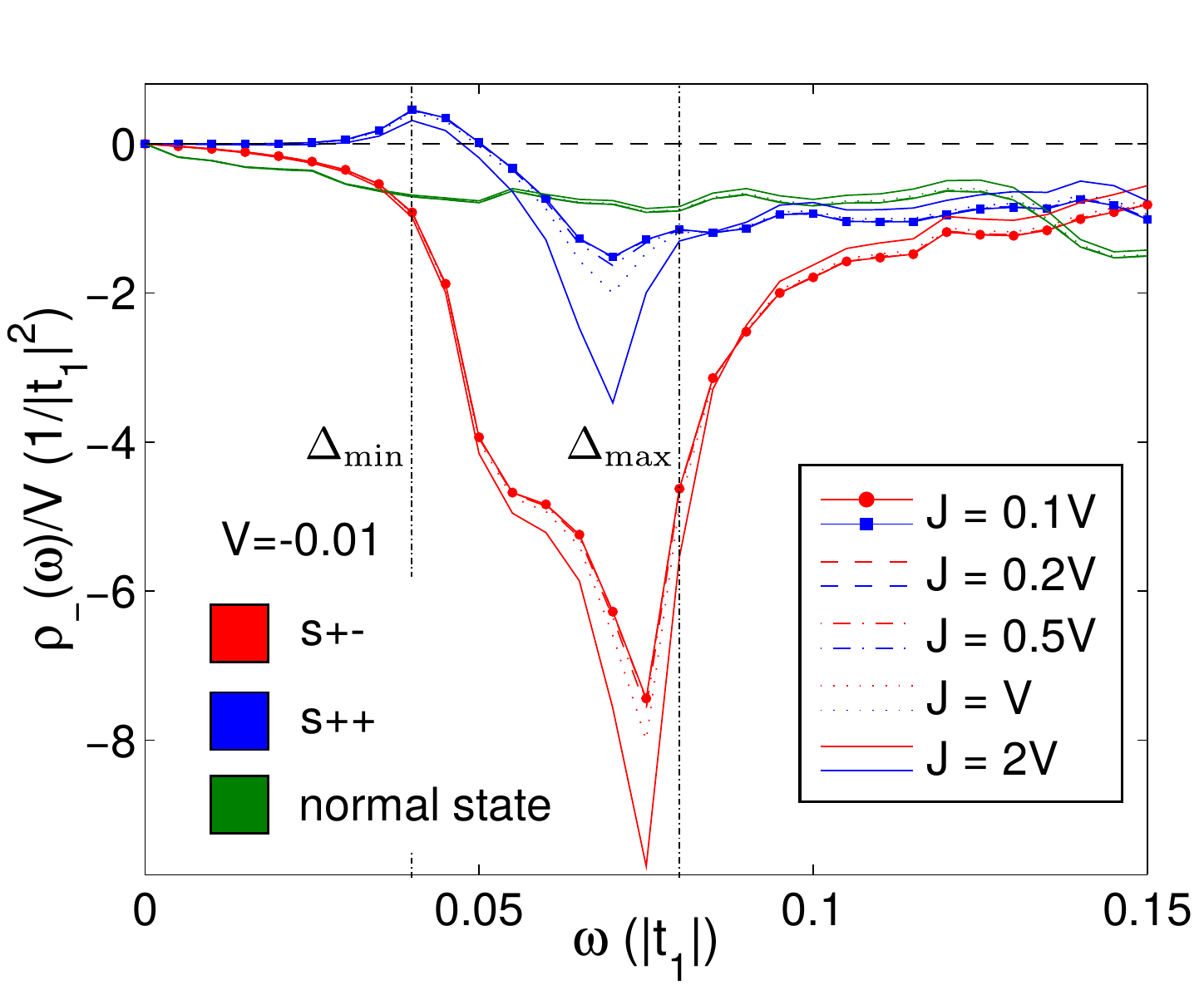}
\caption{
 Antisymmetric QPI signal from T-matrix calculation for a combination of potential scatterer with $V=-0.01|t_1|$ and additional magnetic scatterer with \mbox{$J=\{0.1, 0.2, 0.5, 1, 2\}V$}. In this case the spin summed result as expected from a non-spinpolarized STM experiment is shown. As argued in Ref. \onlinecite{HAEM_2015},  the contribution of  magnetic scattering to $\delta\rho^-$ vanishes in the Born limit, and is generically small, such that the difference in signal between sign-changing order parameter ($s_{\pm}$ red) and the non-sign changing order parameter ($s_{++}$ blue)  is robust against adding a magnetic scatterer.
\label{fig:AK2}}
\end{figure}

A direct calculation of $\delta \rho_{inter}^{-}(\omega)$ from the T-matrix expression with a k-mesh of $500\times500$ using the same band structure and SC mean fields is found to yield nearly identical results, as illustrated in Fig.~\ref{fig:AK1}.
At the same time, changes in the sign or the magnitude of the impurity potential do not alter the qualitative results. We also checked that this result is robust against adding a magnetic scattering contribution (see Eq. (\ref{eq_mag_V})). As shown in Fig. \ref{fig:AK2}, the QPI signal $\delta \rho_{inter}^{-}(\omega)$ is strong and of the same sign in the range $\Delta_{\text{min}}<\omega<\Delta_{\text{max}}$ for $s_\pm$ and changes sign in this energy range for $s_{++}$.\\

\begin{figure}[tb]
\centering

\subfigure{
\includegraphics[width=1\linewidth]{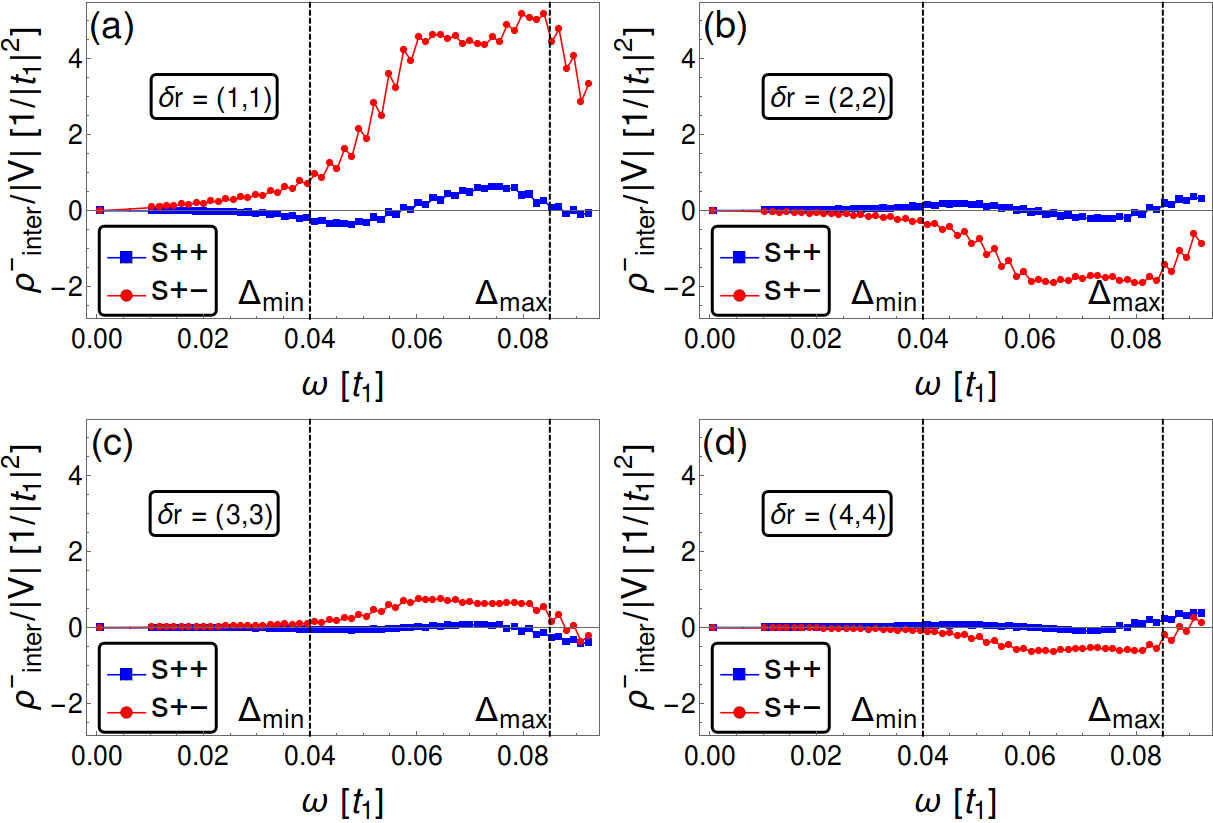}} 
\caption{\fl{a-d} $\delta \rho^-_{inter}(\omega)$ versus $\omega$ for a single impurity displaced diagonally from the central position of the lattice, $\vec{\delta r}= (|\delta r|,|\delta r|$), with $\delta r =1,2,3,4$. As discussed in the text, the curves change overall sign as $(-1)^{|\delta r|}$ and decay with the displacement distance. }

\label{fig:ImpMoved}
\end{figure}

Having reproduced numerically the single central impurity result from Ref.~\onlinecite{HAEM_2015}, we next explore how this result is altered in experimentally more realistic conditions. Initially, however, focusing still on a single impurity, it is notable that the curves in Fig.~\ref{fig:RhoInter1} are sensitive to the centering of the impurity in the FoV. For example, moving the impurity off-center yields very different curves, as demonstrated for diagonal lattice displacements in Fig.~\ref{fig:ImpMoved}. These results follow directly from the definition of the Fourier transform, i.e. a displacement $\vec{\delta r}$ from the central site yields a phase factor $e^{-i\vec{q}\cdot \vec{\delta r}}$, and since we sample the interband signal at $\vec{q} \approx (\pm \pi,0), (0,\pm \pi)$ for the current band, we roughly reproduce the curve with the centered impurity modulo an overall sign change at each lattice step. As seen, the curves are found to degrade as the center moves further out, which follows since the error in the above approximation (i.e. $\vec{q}$ not being exactly $(\pm \pi,0), (0,\pm \pi)$) scales with the displacement distance.\\

\begin{figure}[tb]
\centering
\subfigure{
\includegraphics[width= 1\linewidth]{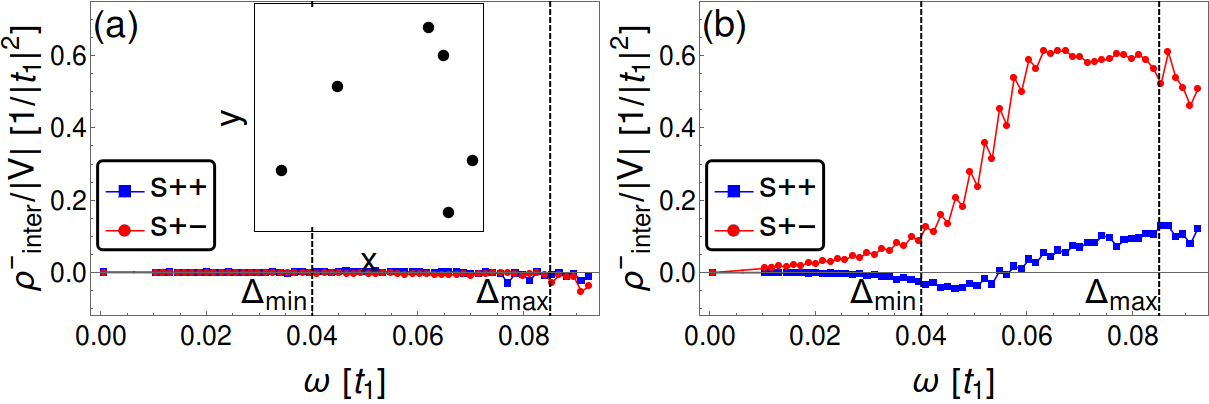}} 
\caption{\fl{a} $\delta \rho^-_{inter}(\omega)$ versus $\omega$ for a system with $0.05\%$ randomly positioned impurities (inset), demonstrating the destruction of the central impurity result from Fig.~\ref{fig:RhoInter1}. Note that the impurities are point-like. \fl{b} The result of dividing out the effective multi-impurity potential. This returns the central impurity curves for $\delta \rho^-_{inter}(\omega)$, with clear distinction between the $s^{++}$ and $s^{\pm}$ gap structures.}
\label{fig:6Imp}
\end{figure}

In the presence of a finite  amount of dilute disorder, we find that $\delta \rho_{inter}^{-}(\omega)$ is generally strongly modified, and it is initially not possible to distinguish  $s_\pm$ from $s_{++}$. For example, in the case of a finite impurity concentration with $0.05\%$ randomly positioned impurities in the $101 \times 101$ lattice Fig.~\ref{fig:6Imp} \fl{a} demonstrates the destruction of the central impurity result from Fig.~\ref{fig:RhoInter1}. In such a system it is clear that further analysis is required to distinguish between different structures. \\
We now show how the central impurity curves for $\delta \rho^-_{inter}(\omega)$ can be recovered, thereby restoring a differentiation between $s_{++}$ and $s_{\pm}$, also for these realistic dilute impurity configurations. Three methods are investigated in the following: 1) Correcting the FT-LDOS signal by the effective multi-impurity potential, 2) choosing a FoV centered at a single well-separated impurity, and 3) moving to a FoV with a much larger impurity concentration where the multi-impurity potential self-averages, thereby essentially restoring the result from the single centered impurity shown in Fig.~\ref{fig:RhoInter1}. \\

In the presence of dilute off-center impurities at sites $\vec{R_j}$, the point-like central impurity potential is replaced by the total impurity potential
\begin{align}
V_{tot}(\vec{q})  &= V_{\text{imp}}\sum_j e^{-i\vec{q}\cdot \vec{R_j}}.
\end{align}
It is then evident  that the single centered impurity result for $\delta \rho^-_{inter}(\omega)$ can be recovered, for sufficiently weak impurities, by simply dividing out this potential in the FT-LDOS\cite{Capriotti03}. The result of which is shown in Fig.~\ref{fig:6Imp}\fl{b}, clearly restoring the central impurity HAEM curves. Dividing out the impurity potential should always return the central impurity  result, yet the exact form of $V_{tot}(\vec{q})$ will in general be very hard to establish experimentally. 

\begin{figure}[tb]
\centering

\subfigure{
\includegraphics[width=1\linewidth]{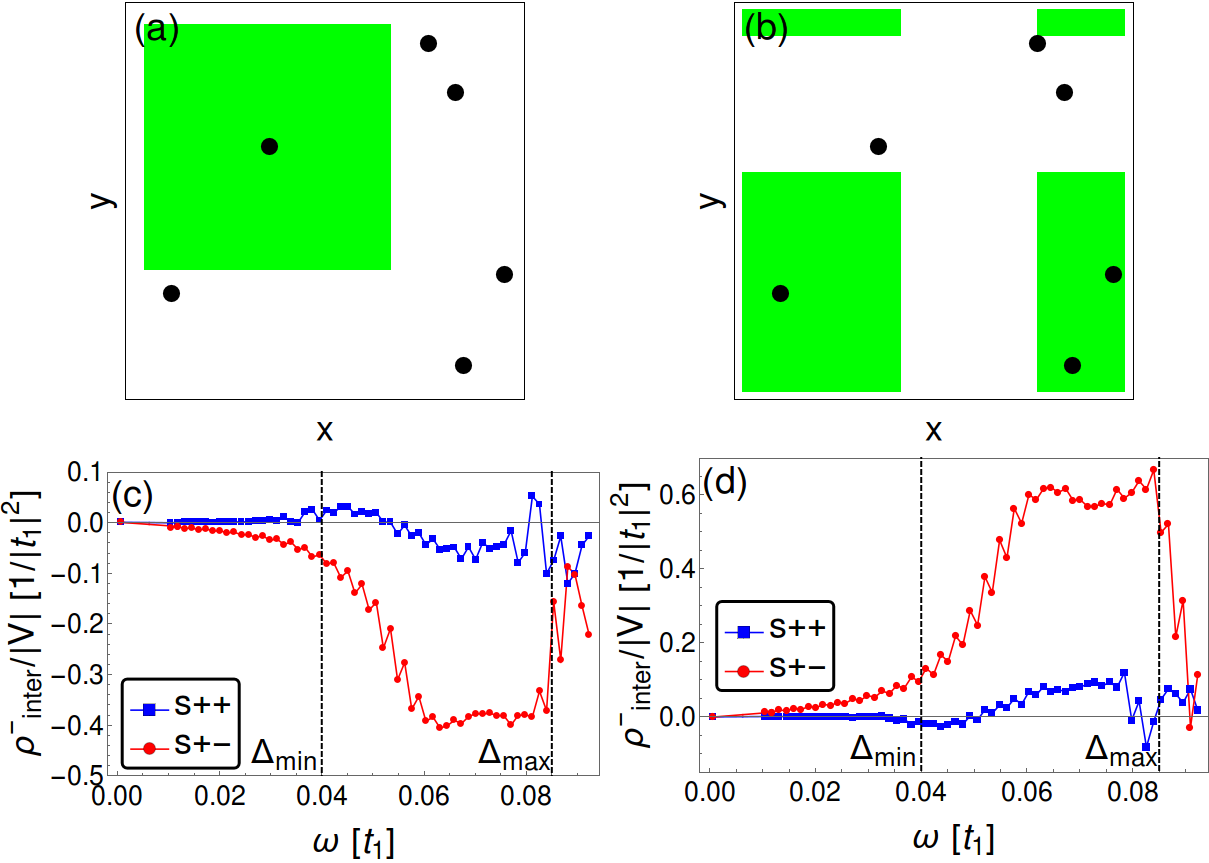}} 

\caption{\fl{a-b} Two different choices of the field of view (green) in the previously introduced dilute impurity system. Each FoV is centered on a well-separated impurity. \fl{c-d}  $\delta \rho^-_{inter}(\omega)$ versus $\omega$ computed for the FoV in \fl{a,b}, respectively. The curves become stable with FoV dimension at ($L_{FoV} \sim 60$ sites), and the result from a single center impurity is recovered.}

\label{fig:FoV}
\end{figure}

In Fig.~\ref{fig:FoV} we consider instead the technique of isolating a well-separated impurity and computing the FT centered on this impurity site with a reduced FoV, similar to the method utilized in Ref.~\onlinecite{OSP1}. The dilute impurity system considered so far enables this analysis for multiple impurities. Choosing a suitable impurity, we compute the $\delta \rho^-_{inter}(\omega)$ curves by this method starting from a small FoV dimension $L_{FoV}$. The signal is initially noise dominated (not shown), but converges to the central impurity $\delta \rho^-_{inter}(\omega)$ result as $L_{FoV}$ is increased. Fig.~\ref{fig:FoV} \fl{c-d} displays these curves computed past the convergence point for two different well-isolated impurities \fl{a-b} in the same configuration as Fig. \ref{fig:6Imp}, with a large window dimension $L_{FoV} =65$a. As seen, we again recover the ability to differentiate between sign-changing and sign-preserving gap structures. We have confirmed this result for all impurities included in the configuration displayed, as well as for impurities in different dilute impurity configurations.  \\

\begin{figure}[tb]
\centering

\subfigure{
\includegraphics[width=1\linewidth]{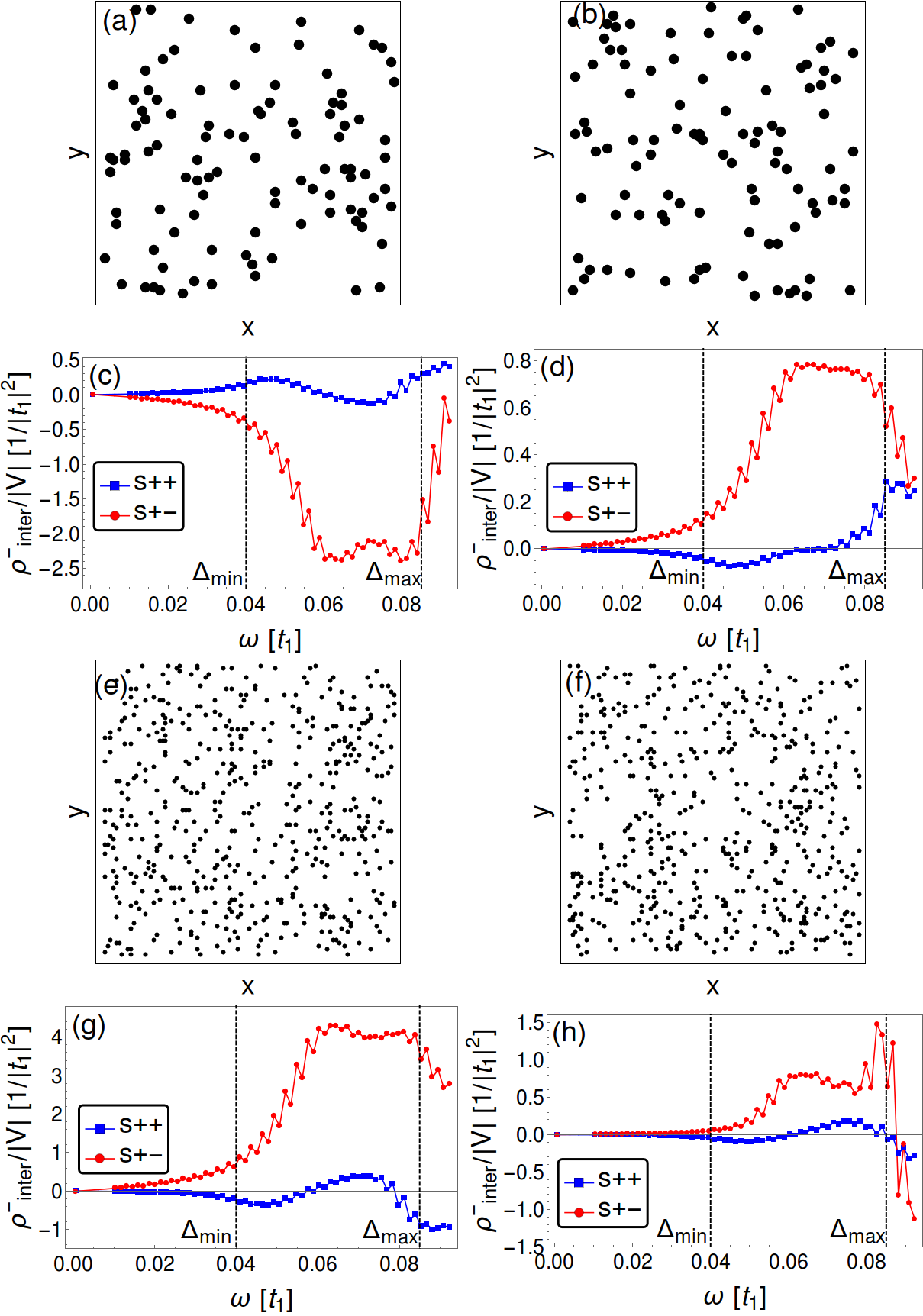}} 

\caption{\fl{a-b} Two different $1\%$ impurity configurations alongside \fl{c-d} $\delta \rho^-_{inter}(\omega)$ versus $\omega$ for these systems. The impurity potential for this large impurity concentration self-averages, and the central impurity HAEM result is recovered. \fl{e-h} Similar calculation for two different $5 \%$ impurity (point-like) configurations, where the self-averaging is seen to be retained. Note that the impurities are again point-like and the black dots have different sizes purely for plotting purposes.
}

\label{fig:ManyImp}
\end{figure}

Finally, we will demonstrate the robustness of the HAEM method in the large impurity concentration (dirty) limit. These systems preclude the selection of a well-isolated impurity, and thus require separate consideration. We have found that the effective impurity potential self-averages in the systems, resulting in the reproduction of the central impurity HAEM result for $1\%$ and $5 \%$ point-like impurities with no need for the above procedures. This result is shown in Fig.~\ref{fig:ManyImp}, where the dense impurity configurations produces easily distinguishable HAEM curves in multiple distinct configurations. 

\begin{figure}[tb]
\includegraphics[width= \linewidth]{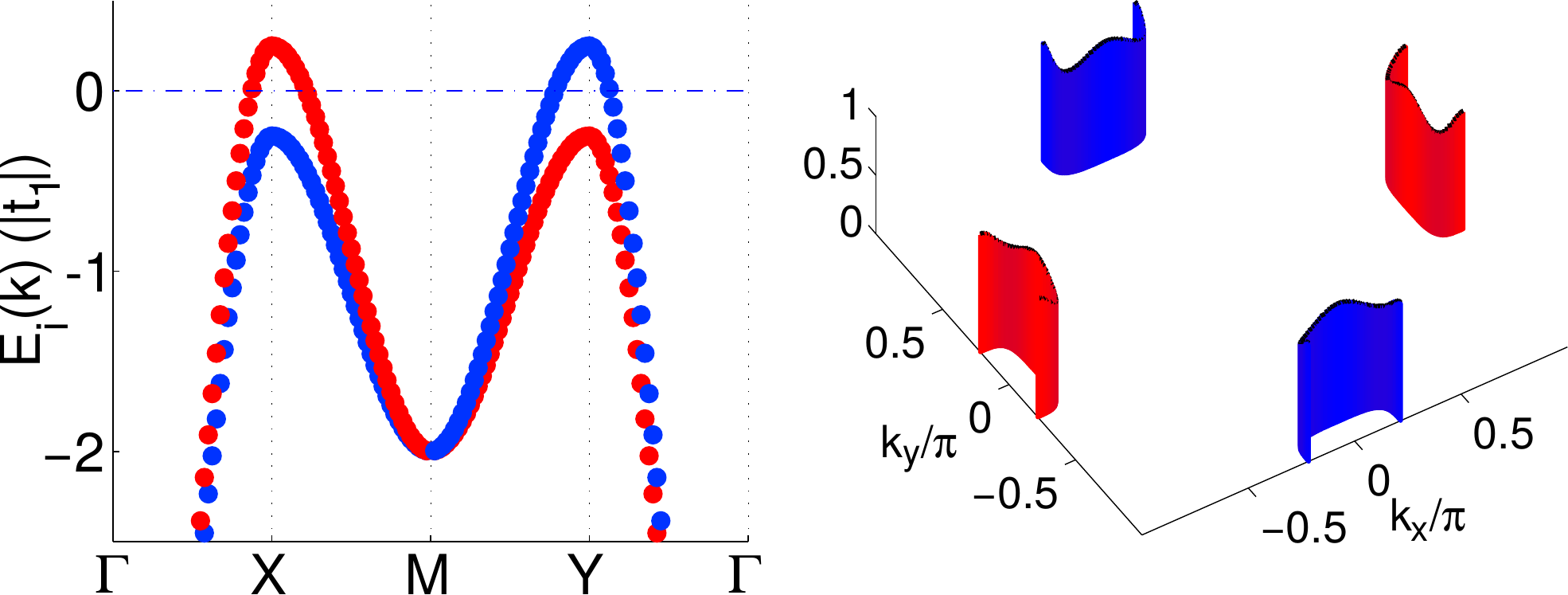}
 \rput[tr](-0.45\linewidth,0.42\linewidth){(a)}
  \rput[tr](0.06\linewidth,0.42\linewidth){(b)}
 \caption{Band structure along a high symmetry path (a) and Fermi surface (b) of our two orbital model (orbitals indicated by color: red/blue) having only pockets around the X and Y point. The (small) variations of the orbital weight which are necessary to allow for inter-pocket scattering with a orbital-diagonal impurity potential, are plotted as the third direction (b).}
 \label{fig_two_band_d}
\end{figure}

\begin{figure}[tb]
\includegraphics[width= \linewidth]{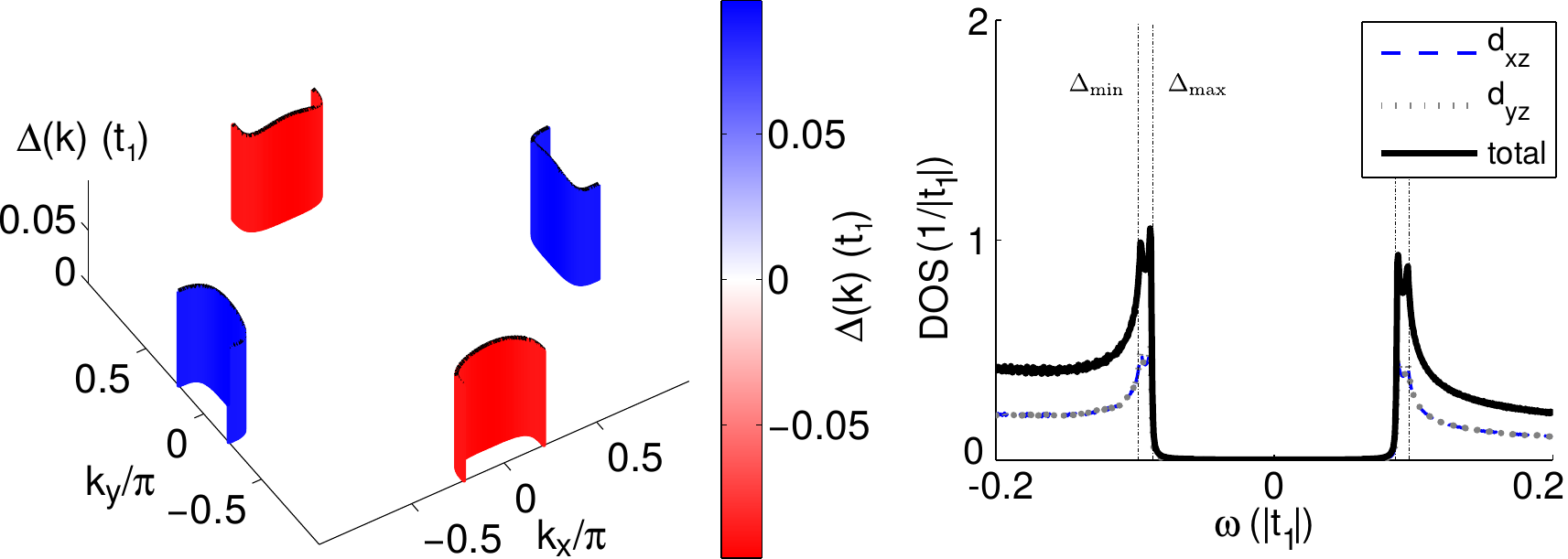}
 \rput[tr](-0.45\linewidth,0.4\linewidth){(a)}
  \rput[tr](0.1\linewidth,0.4\linewidth){(b)}
  \caption{Order parameter for our $d$-wave calculation (a) together with the density of states showing a small anisotropy apparent by two coherence peaks at very similar energies $\Delta_{\text{min}}=0.088$ and $\Delta_{\text{max}}=0.096$ (b). The order parameter without sign change (not shown) yields exactly the same density of states.}
 \label{fig_two_band_gap_d}
\end{figure}

\begin{figure}[t]
\includegraphics[width= \linewidth]{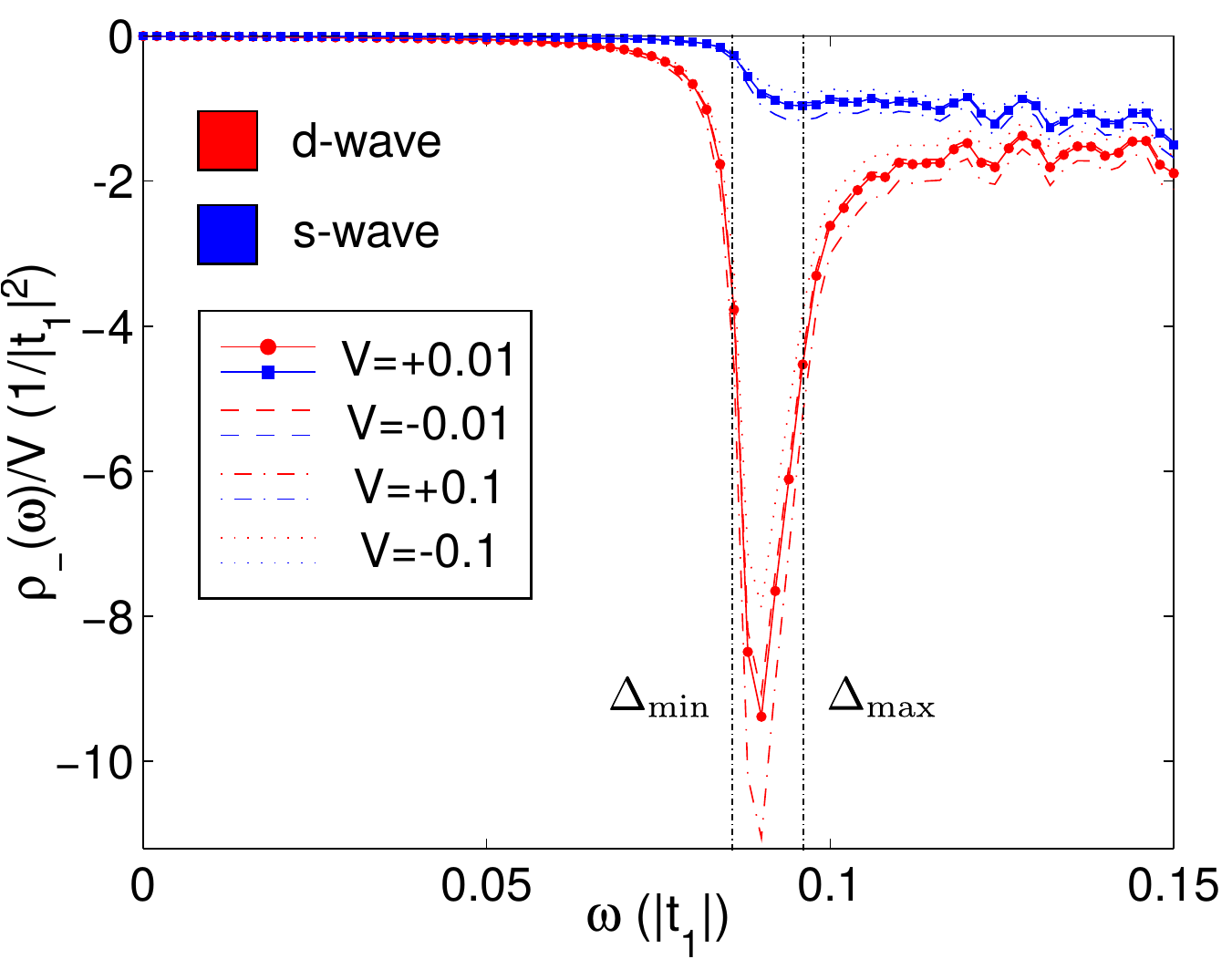}
 \caption{Antisymmetrized QPI $\rho_-(\omega)$ for the nodeless $d$-wave scenario: The sign-changing order parameter produces a sharp peak at $\Delta_{\text{min}}<\omega<\Delta_{\text{max}}$ (red curves) while this signal is absent for the calculation with the same gap magnitude everywhere, but without sign change (blue curves). The curves are divided by the sign of the impurity potential to show that  negative and positive potentials yield (apart from the trivial difference in the sign) the same qualitative behavior (dashed, solid lines).}
 \label{fig_rho_minus_d}
\end{figure}

\subsection{d-wave}
In this section, we show explicitly what the HAEM scheme expects to find for a fully gapped $d$-wave state as might be a candidate for the ground state of one of the Fe-chalcogenide systems with electron pockets only at the Fermi level\cite{HirschfeldCRAS}. For this purpose, we need to modify the Hamiltonian so that  Fermi sheets exist only around the $(\pi,0)$ and $(0,\pi)$.  This will give a  tendency towards a $d_{x^2-y^2}$-wave instability in the superconducting channel but will  avoid a nodal state in the absence of a $\Gamma$-centered Fermi surface. 
We therefore choose $\tilde t_1=0.5625t_1$, $\tilde t_2=0.4375 t_1$, $\tilde t_3=0.5t_1$, and $\tilde t4=0.300 t_1$ to yield the band structure shown in Fig. \ref{fig_two_band_d}(a) and the Fermi surface together with the orbital content as shown in Fig. \ref{fig_two_band_d}(b). The $d$-wave order parameter consisting of the lowest harmonic only, which we use as our test order parameter, displays small anisotropy over the Fermi surface as seen in Fig. \ref{fig_two_band_gap_d} with $\Delta_{\text{min}}\approx 0.088 t_1$ and $\Delta_{\text{max}}\approx 0.096 t_1$. Within a non-selfconsistent T-matrix calculation\cite{OSP1} we can implement the sign removal exactly, i.e. obtain two models which have exactly the same DOS as presented in Fig.\ref{fig_two_band_gap_d}(b). Instead of calculating the LDOS in real space and then Fourier transforming, we work in momentum space, set up the Greens functions and calculate the T-matrix with a constant scattering potential describing non-magnetic scatterers as expected for impurities centered at the lattice positions. In contrast to the case discussed in the previous section, the scattering processes involving sign-changes in the $d$-wave case are around $(\pi,\pi)$, such that we integrate over one quarter of the Brillouin zone to construct the antisymmetrized QPI signal, Eq.(\ref{eq_anti_qpi}). As expected from the  small anisotropy, the sign-changing order parameter produces a sharp peak close to the gap energy, as seen in Fig.\ref{fig_rho_minus_d} which is completely absent in the calculation of the same quantity for the model without sign-change. Noting that this result is robust against changes in the magnitude of the impurity potential, it could be used to test the two different scenarios without too many assumptions from the theoretical modeling.

\section{Conclusions}

We have analyzed the robustness of the proposed QPI procedure for distinguishing between sign-preserving and sign-changing gap structures in multi-band superconductors\cite{HAEM_2015}. The field of view centered single-impurity result for the momentum-integrated antisymmetrized interband QPI response changes qualitatively for off-center positions and/or in the presence of dilute concentrations of disorder. We have shown, however, three possible ways to recover the robust determination of inter-pocket gap sign changes, and conclude therefore that the HAEM procedure is a realistic method to be used for this purpose. 

\section{Acknowledgements}

The authors acknowledge useful discussions with J.C. Davis, I. Eremin, A. Kostin, I.I. Mazin,  P. Sprau,  H.-H. Wen, B.M.A.\ acknowledges support from Lundbeckfond fellowship (grant A9318).  P.J.H. was supported by NSF-DMR-1407502.

\end{document}